\documentstyle[amssymb,aps,multicol]{revtex}

\begin{document}
\preprint{}
\draft

\title{Universal Control of Decoupled Quantum Systems}
\author{Lorenza Viola${}^{1}$, Seth Lloyd${}^1$, and Emanuel
Knill${}^{2\,\dagger}$  
% \thanks{ Electronic addresses: vlorenza@mit.edu; slloyd@mit.edu;
% knill@lanl.gov }
}
\address{ ${}^1$ d'Arbeloff Laboratory for Information Systems and 
Technology, %\\
Department of Mechanical Engineering, \\ Massachusetts Institute of 
Technology, %\\
Cambridge, Massachusetts 02139 \\
${}^2$ Los Alamos National Laboratory, Los Alamos, New Mexico 87545} 
\maketitle

\begin{abstract}
It is shown that if one can perform a restricted set of fast
manipulations on a quantum system, one can implement a large class of
dynamical evolutions by effectively removing or introducing selected 
Hamiltonians. The procedure can be used to achieve universal 
noise-tolerant control based on purely unitary open-loop 
transformations of the dynamics. As a result, it is in principle 
possible to perform noise-protected universal quantum computation 
using no extra space resources.
\end{abstract}


% 03.65.-w fundamental quantum mechanics
% 03.65.Bz Foundations, theory of measurement, miscellaneous
% 03.67.-a quantum information 
% 03.67.Lx quantum computation
% 89.70.+c Information science 

\begin{multicols}{2}

The desire to shape quantum evolution according to precisely
controlled dynamics is shared by various areas of contemporary physics
and engineering \cite{blaquierie}.  A problem commonly encountered in
the task of controlling the dynamical behavior of a quantum system is
the need for removing unwanted interactions present in the full
Hamiltonian: historically, one of the first solutions was provided by
Nuclear Magnetic Resonance spectroscopy, where a variety of {\sl
decoupling} techniques have been developed to simplify spectra by
effectively eliminating selected contributions to the nuclear
Hamiltonian \cite{ernst}.  Once suppression of a given term is
obtained, the corresponding Hamiltonian may no longer be directly
available for control.  From the perspective of attaining universal
dynamical control, this raises the question of devising ways for
introducing or re-introducing Hamiltonian control compatible with the
prescribed decoupling action.

A striking example is the case of an open quantum system, where the
coupling to the environment is responsible for inducing quantum
decoherence and dissipation processes, thereby corrupting the original
unitary dynamics \cite{weiss}. 
The demand for universal control strategies of noise-decoupled quantum 
systems, able to effectively reject environmental noise while still ensuring 
full control capabilities, has been heightened tremendously due to the 
challenge of implementing quantum computation \cite{divincenzo}. 
In spite of the beautiful and powerful advancements made in the theory 
of quantum error correction \cite{qecc}, fault-tolerant error correction 
\cite{fault} and concatenated coding \cite{laflamme}, 
practical exploitation of these results is still seriously constrained 
by the amount of extra space resources required \cite{nature}.

In this Letter, we address the general problem of {\sl open-loop}
controllability of a decoupled quantum system: the controller is 
assumed to apply time-dependent potentials without ever measuring 
the actual state of the system.
We introduce {\sl programming} procedures for combining the desired 
control action with the decoupling operations and identify the conditions 
under which universal control over the effective decoupled dynamics is
retained.  
As a consequence, we demonstrate the possibility of achieving
noise-tolerant control of open quantum systems solely based on unitary
manipulations which do not require ancillary resources.

{\it Decoupling.$-$} We begin by recalling the essential ingredients
of decoupling in the language of \cite{viola2}. Let $H$ be the
Hamiltonian of a quantum system living in a Hilbert space ${\cal
H}$. Suppose we have the capability of performing instantaneously a
certain set of unitaries, meaning that the corresponding set of
Hamiltonians can be turned on for negligible amounts of time $\tau$
with (ideally) arbitrarily large strength.  We shall term such
impulsive full-power control operations as {\sl bang-bang} (b.b.)
controls \cite{viola2,viola1}.  If $U$ is a rotation that can be
implemented b.b., it is conceivable that $U^{-1}=U^\dagger$ can be
realized b.b. as well.  In this case, the set of realizable
b.b. operations is a subgroup ${\cal G}_{b.b.}$ of the full group
${\cal U}({\cal H})$ of unitary transformations over ${\cal H}$. A
{\sl decoupler} on ${\cal H}$ operates by iterating the system through
a cyclic time evolution which judiciously combines sequences of
b.b. operations with free evolutions under the natural Hamiltonian
$H$.  Accordingly, a decoupler shall be characterized by a finite
group ${\cal G}\subseteq {\cal G}_{b.b}$ of b.b. operations ({\sl
decoupling group}) together with a known time scale $T_c$ ({\sl cycle
time}) determining the duration of a single cycle.  Full knowledge of
the decoupler operations, including the exact sequence of group
operations $g_j$, $j=0, \ldots, |{\cal G}|-1$, 
$|{\cal G}|\equiv \text{ord}({\cal G})$, and their temporal
separation $\Delta t$, may not be available from the beginning
(black-box decoupler).

What does a decoupled evolution look like ? 
A convenient picture is provided by average Hamiltonian theory 
\cite{ernst,viola2}. Consider a slicing of a given evolution time $T$ 
as $T=NT_c\equiv N |{\cal G}| \Delta t $,  
and let $U_0(\Delta t)=\exp(-i H \Delta t)$ be the free propagator.
The evolution in the presence of the decoupler is uniquely determined by 
the sequence of group transformations $\{g_j\}$ over a single cycle time:
\begin{equation}
U(T_c)= \prod_{j=0}^{|{\cal G|}-1} \,g_{j}^\dagger U_0(\Delta t) g_j
\equiv e^{-i H_{eff} T_c } \;,
\label{cycle}
\end{equation}
$H_{eff}$ denoting the resulting effective Hamiltonian. In the ideal 
limit of arbitrarily fast cycle time $T_c \rightarrow 0$, with 
$N \rightarrow \infty$ in such a way that $NT_c=T$, $H_{eff}$ approaches
\begin{equation}
H \mapsto H_{eff} = {1 \over |{\cal G}| } 
\sum_{g_j \in {\cal G}} \, g_j^\dagger\,  H \, g_j 
\equiv \Pi_{\cal G}(H) \;.
\label{projection}
\end{equation}
Eq. (\ref{projection}) defines a quantum operation on the space
End(${\cal H}$) of bounded operators on ${\cal H}$. Physically,
$H_{eff}$ represents the leading contribution to the average
Hamiltonian describing the motion of the system under the influence of
the decoupling field, the $j$-th term in the sum (\ref{projection})
being identical to the so-called {\sl toggling frame} Hamiltonian
during the $j$-th cycle subinterval \cite{ernst}.  From a geometric
point of view, the map $\Pi_{\cal G}$ introduced in (\ref{projection})
can be identified with the projector on the so-called {\sl
centralizer} $Z({\cal G})$ in End(${\cal H})$
\cite{viola2,zanardi1}:
\begin{equation}
Z({\cal G}) =  \{ \, {\cal O} \,|\,
[ {\cal O}, g_j]=0 \;\; \forall g_j \in {\cal G} \,\} =
\Pi_{\cal G}(\text{End}({\cal H}))\;. 
\label{commutant}
\end{equation}
$Z({\cal G})$ is a subalgebra of End(${\cal H})$. We shall denote by 
$Z_H({\cal G})$ the subspace of Hermitian operators belonging to 
$Z({\cal G})$. Eq. (\ref{commutant}) allows a direct interpretation of the 
decoupler action in terms of symmetry properties: since 
$H_{eff}=\Pi_{\cal G}(H)$, the decoupled evolution is {\sl symmetrized} 
according to the group ${\cal G}$ \cite{zanardi1}. 
All the components of the dynamics generated by $H$, which are {\sl not} 
invariant under the group ${\cal G}$, are effectively averaged out. 

The group-theoretical prescription (\ref{cycle}) translates directly
to pulse control, a control cycle involving a sequence of decoupling
pulses $D_j=g_j g_{j-1}^\dagger$, $j=1, \ldots,|{\cal G}|$, separated
by delays of free evolution $\Delta t$ and fulfilling $D_{|{\cal
G}|}\ldots D_2 D_1 =\openone$ by cyclicity. The limit $T_c \rightarrow
0$ cannot be met exactly in practice.  While symmetry properties
suffice to specify the output of a given decoupler, time scales are
crucial in determining how good the decoupler performs in a realistic
scenario where both the pulse duration $\tau$ and the cycle time $T_c$
are finite. Qualitatively speaking, the projection on the centralizer
(\ref{projection}) will tend to {\sl symmetrize interactions whose
typical correlation times are long on the time scale determined by
$T_c$}. If $\tau_c$ represents the shortest correlation time
associated with the unwanted interactions, decoupling will be
effective under the hierarchy of time scales $ \tau \ll {\Delta t}
\leq T_c \ll \tau_c $ \cite{viola2}.  Effects due to finite
pulse duration are expected to be of order O$(\tau/\tau_c)$, while
cycle time corrections scale with a ratio O($T_c/\tau_c)$
\cite{viola2}. It is worth stressing that $T_c$ determines the
minimum time scale over which decoupling is guaranteed. Thus, the
original continuous-time dynamics under $H$ is replaced by a {\sl
stroboscopic} time development under $H_{eff}$ 
with a natural time unit equal to $T_c$.

For open quantum systems, the above description comprises two coupled
subsystems $S$ and $B$ with associated Hilbert space 
${\cal H} = {\cal H}_S \otimes {\cal H}_B$, the $S$-component 
representing the system of physical interest ({\it e.g.}, the computational 
degrees of freedom of a quantum computer). If $H=H_S +H_B+H_{SB}$ denotes 
the total Hamiltonian, environmental noise is introduced through a set of
(traceless) error operators $E_\alpha$ in the interaction Hamiltonian,
$H_{SB}=\sum_\alpha \, E_\alpha \otimes B_\alpha$, $B_\alpha$ denoting
bath operators.  We require the linear space ${\cal E}=
\text{span}\{E_\alpha\}$ to be finite dimensional. An error space
${\cal E}$ will be {\sl correctable} by any decoupler ensuring that
$\Pi_{\cal G}({\cal E})= 0$. The appropriate correlation time $\tau_c$
is related to the memory time of the reservoir $B$ \cite{viola2}. Note
that this necessarily implies a non-Markovian error scenario.
  
The capability of successfully decoupling the system $S$ from the
effects of the environment is only a first step to achieving
noise-tolerant control of $S$-degrees of freedom. Can we do better?

{\it Programming.$-$} The presence of a decoupler imposes restrictions
on the manipulations available to further control the system
dynamics. Those restrictions are in the form of both {\sl symmetry}
and {\sl timing} constraints. Suppose, for instance, that a
Hamiltonian $A$ is applied over a time interval long compared to
$T_c$. Then the evolution generated by $A$ may be quenched completely
if $A \not \in Z_H({\cal G})$. If, instead, the action corresponding
to $A$ is implemented in the form of b.b. rotations, it is easy to
check that, unless the insertion points are chosen carefully, any
extra pulse interferes with the decoupler operations, thereby spoiling
noise averaging. This can be avoided, for instance, by making sure
that the b.b. operations are inserted {\sl in between} cycles only.
In general, indiscriminate application of control operations will not
lead to the expected result.  The goal of programming is to characterize 
the degree of control attainable compatibly with the desired decoupling 
action. This is influenced by three factors: available knowledge of the 
error space ${\cal E}$; available knowledge of the decoupler operations; 
available control resources. In particular, since switching on/off strong
interactions for short times is difficult in practice, a relevant criterion 
will be trying to keep the number of required b.b. operations to a minimum.

Let us first consider the case where the error space ${\cal E}$ is known
and a decoupling group ${\cal G}$ exists with a nontrivial centralizer,
$Z({\cal G}) \not = \{\openone\}$, a situation corresponding to
selective averaging according to \cite{viola2}. By virtue of 
(\ref{projection}), it is always possible to apply slowly any Hamiltonian 
$A \in Z_H({\cal G})$ in parallel with the decoupler. However, since the 
unitary evolutions generated by such Hamiltonians also lie by construction in
the centralizer of ${\cal G}$, universality on the full Hilbert space
cannot be achieved by purely exploiting this kind of {\sl weak
(strength)/ slow (switching) control}. Supplementary coding methods
are demanded, which shall be discussed elsewhere \cite{next}. Instead,
we focus here on the idea of combining slow control from the set
$Z_H({\cal G})$ with suitable fast manipulations available in addition
to the decoupling ones.

A relatively straightforward situation occurs when the group ${\cal
G}_{b.b.}$ of realizable b.b. operations is large enough to accommodate
{\sl two} distinct decouplers with known (possibly different) cycle
times, {\it i.e.}, $\Pi_{{\cal G}}({\cal E})=0$, $\Pi_{\tilde{\cal
G}}({\cal E})=0$, with ${\cal G}, \tilde{\cal G} \subset {\cal
G}_{b.b.}$. Notice that if ${\cal G}$ and $P$ are respectively a
decoupling group and a unitary transformation, $P \not \in {\cal G}$,
$P \not \in Z({\cal G})$, then $\tilde{\cal G}=P^\dagger {\cal G} P$
implements a {\sl twisted decoupler} of the same order provided
$P^\dagger {\cal E} P ={\cal E}$. Suppose now we can apply
Hamiltonian $A \in Z_H({\cal G})$ for a time interval $\Delta T_1 =
N_1 T_{c1}$ (in parallel with decoupler 1), Hamiltonian $B\in
Z_H(\tilde{\cal G})$ for a time interval $\Delta T_2 = N_2 T_{c2}$ (in
parallel with decoupler 2), etc., ({\it e.g.}, one could have
$A=\Pi_{{\cal G}}(H)$, $B=\Pi_{\tilde{\cal G}}(H)$). Then, by using
standard universality results \cite{seth}, any $U=e^{L}$ could be
created, where $L$ belongs to the Lie algebra generated by $iA,iB$
under commutation.  Accordingly, universal control over the decoupled
dynamics is achieved whenever this algebra amounts to the whole Lie
algebra of anti-Hermitian operators.

Even if only a single decoupler ${\cal G}$ is available, a similar
strategy can be mimicked through a simple trick. Suppose that, in
addition to decoupling pulses in ${\cal G}$, we can perform on the
system a b.b. rotation $P \in {\cal G}_{b.b.}$. What can we do with
this capability ?  Assuming that we are able to {\sl synchronize}
operations with the cycle time, we can make the system effectively
evolve according to a {\sl transformed} average Hamiltonian. Let
$\Delta T=NT_c$ be a given time window, with decoupled evolution ruled
by the Hamiltonian (\ref{projection}).  Imagine now inserting a pulse
$P$ immediately before the beginning of $\Delta T$, followed by a
pulse $P^\dagger$ synchronized with the end of $\Delta T$. Then
evolution over $\Delta T$ can be described in terms of a new average
Hamiltonian $\tilde{H}_{eff}=P^\dagger \Pi_{\cal G}(H) P$ {\it i.e.},
\begin{equation}
\tilde{H}_{eff} = {1 \over |\tilde{\cal G}| } 
\sum_{\tilde{g}_j \in \tilde{\cal G}} \, \tilde{g}_j^\dagger\, \tilde{ H} 
\, \tilde{g}_j \equiv \Pi_{\tilde{\cal G}}(\tilde{H}) \;,
\label{projectiontilde}
\end{equation}
where $\tilde{H}=P^\dagger H P$ and a twisted decoupling group
$\tilde{\cal G}=P^\dagger {\cal G} P$ has been defined, with
associated centralizer $Z(\tilde{\cal G})=P^\dagger Z({\cal G})P$. 
Thus, the net effect of the two programming pulses $P,P^\dagger$
amounts to implement decoupling according to $\tilde{\cal G}$, noise
averaging being retained since {\sl both} the original decoupling
group ${\cal G}$ and the error space ${\cal E}$ are simultaneously
rotated. Clearly, one has to ensure that $P \not \in Z({\cal G})$ in
order to steer the effective Hamiltonian out of the original
centralizer $ Z({\cal G})$.

Let $A$ now be, as above, a realizable Hamiltonian in the centralizer
of ${\cal G}$ and let $B=P^\dagger A P$ denote its rotated counterpart.  
Then, by alternating evolution periods according to $A$ and $B$, the 
latter being obtained by inserting pairs of pulses $P,P^\dagger$ with 
appropriate timing, it is possible in principle to obtain any Hamiltonian 
in the algebra generated by $A,B$ under commutation.  The reasoning is 
easily extended to the case where a given choice of interactions is 
realizable in $Z_H({\cal G})$. In practice, the advantageous feature of 
this scheme is that by performing a single extra b.b. operation, a new 
repertoire of Hamiltonians becomes effectively available for slow control 
in the centralizer $Z(\tilde{\cal G})$.  In the generic case, under the
conditions given in \cite{seth}, any desired unitary transformation
will be reachable in principle, thereby implying complete control of
the decoupled evolution.

Note that knowledge of the exact decoupling sequence has not been
exploited so far, implying validity of the previous schemes even for
black-box decouplers.  If detailed information on the decoupler
operations is available, this knowledge can be used to devise
alternate control schemes implying less stringent resources.  Suppose
that we want to reintroduce control  by some Hamiltonian $B
\not \in Z_H({\cal G})$. If $e^{iB}\in {\cal G}_{b.b.}$, one could
always, in principle, exploit the freedom of inserting such a
b.b. pulse at the beginning and/or the end of decoupling cycles
without affecting decoupling itself.  Actually, it is possible to
replace the the b.b. requirement with a weaker assumption, by
imagining that the strength of $B$ cannot be made arbitrarily large
but $B$ can still be turned on and off arbitrarily fast. In other
words, let us assume a form of {\sl weak (strength)/ fast (switching)
control} whereby Hamiltonians can be modulated at the same rate as the
b.b. control within a cycle.  Then to reintroduce control according to
$B$ over a time interval $\Delta T=NT_c$, it suffices to turn it on
during the $\openone$-frame subinterval of each decoupling cycle. The
evolution is ruled by the effective Hamiltonian
\begin{equation}
\tilde{H}_{eff} ={1 \over |{\cal G}| } \bigg(
\sum_{g_j \in {\cal G}} \, g_j^\dagger\,  H \, g_j + B \bigg) =
\Pi_{\cal G}(H) + \frac{1}{|{\cal G}|}\, B \;, 
\label{drift}
\end{equation}
which acquires a component along $B \not \in Z_H({\cal G})$. Strength
reduction for such a Hamiltonian can be avoided if an enlarged set of
interactions is amenable of fast switching: one just turns on a
Hamiltonian $B_j=g_j B g_j^\dagger$ during the $j$-th subinterval of
each cycle, the overall effect being elimination of the $|{\cal
G}|^{-1}$-factor in front of $B$.

Using the above methods, controlled evolutions can be designed by both
letting the system evolve under the action of the decoupler alone and
by incorporating modified decoupling cycles to displace the effective
Hamiltonian out of $Z({\cal G})$ as in (\ref{drift}). The issue of
complete control can now be addressed by looking at the combined
repertoire of interactions available for slow control in the
centralizer, {\it e.g.}, $A =\Pi_{\cal G} (H) \in Z_H ({\cal G})$,
together with the ones capable of supporting fast modulation, {\it
e.g.}, Hamiltonian $B$ considered above.  Again, the conditions
established in \cite{seth} provide a necessary and sufficient
criterion for universality.

Let us briefly comment on the situation where no knowledge is
available on the error space ${\cal E}$. In this case, decoupling can
be achieved only by maximal averaging \cite{viola2}, so that the
effective Hamiltonian is a trivial $c$-number, $H_{eff}=\lambda\,\openone$. 
Since for decoupling groups with this property $Z({\cal G}) = \{\openone\}$, 
control schemes based on multiple or twisted decouplers are not useful 
anymore.  In principle, one could still attain complete control in two 
circumstances: either ${\cal G}_{b.b.}$ contains a universal set of 
operations, which have to be performed synchronously with the decoupler 
clock $T_c$; or a universal set of Hamiltonians can be switched fast. Even 
with this option, the minimum number of required b.b. operations, $|{\cal
G}|=(\text{dim}({\cal H}_S))^2$ \cite{viola2}, may be very large for relevant 
systems, strength losses in effective Hamiltonians (\ref{drift}) becoming 
possibly quite significant.  We turn now to analyze more specifically the 
case of quantum computation.

{\it Universal Quantum Computation.$-$} Consider a quantum computer
made of $K$ qubits, ${\cal H}_S \simeq ({\sf C}^2)^{\otimes K}$, and
assume that the relevant coupling to the environment can be accounted
by a {\sl linear} interaction of the form
\begin{equation}
H_{SB}= \sum_{a,i}\, \sigma_a^{(i)} \otimes B_a^{(i)}\;, 
\;\;\; a=x,y,z;\; i=1, \ldots K\;, 
\label{linear}
\end{equation}
for suitable bath operators $B_a^{(i)}$. The above coupling
encompasses various models of interest, with error space spanned by
combinations of single-qubit operators.  Following the classification
of \cite{lidar}, dim$({\cal E})=3K$ for independent decoherence where
$\{E_\alpha\}=\{\sigma_a^{(i)} \}$, while dim$({\cal E})=3$ in the
opposite limit of collective decoherence with global generators
$\{E_\alpha\}=\{\sum_i\, \sigma_a^{(i)} \}$, intermediate situations 
occurring with cluster decoherence. 
Selective decoupling of the quantum register from $H_{SB}$ requires a 
minimum of 3 b.b. operations {\it i.e.,} $|{\cal G}|=4$. A convenient 
choice is ${\cal G}=\{ \openone, \otimes_{i=1}^K \sigma_a^{(i)}\}$, 
in which case the $g_j$'s correspond to collective $\pi$-rotations and 
$Z({\cal G})$ is generated by bilinear interactions of the form 
$\sigma_a^{(i)}\sigma_a^{(j)}$ \cite{viola2,gottesman}.

We sketch now the application of the programming schemes described
above. Suppose that, in addition to the ${\cal G}$-decoupler, we are
equipped with a second decoupler based on a group $\tilde{\cal G}$,
where for instance $\sigma_x^{(i)}$ is interchanged with
$\sigma_z^{(i)}$ and $\sigma^{(j)}_y$ with $\sigma_z^{(j)}$ (say,
$i=1,j=2$). Obviously, $\Pi_{\tilde{\cal G}}({\cal E})=0$.  Then
almost any set of gates $U_A=e^{iAt_A}$, $U_B=e^{iBt_B}$ will be
universal over ${\cal H}_S$, noise-suppression being preserved if
Hamiltonians $A \in Z_H({\cal G})$, $B\in Z_H(\tilde{\cal G})$ are
applied in parallel to the decouplers ${\cal G}$, $\tilde{\cal G}$
respectively. Since the latter groups are connected by a double
$\pi/2$-pulse 
$P=\exp(-i(\pi/4)\sigma_y^{(1)}) \exp(-i(\pi/4)\sigma_x^{(2)} )$, the 
same result can be reached if the realizable group of b.b. operations 
only includes ${\cal G}_{b.b.}=\{{\cal G}, P, P^\dagger\}$, but we have 
the capability of inserting $P,P^\dagger$ pulses synchronized with $T_c$. 
Finally, in the case that full knowledge is available of the decoupling 
sequence, a constructive result can be given. Note that the centralizer
$Z_H({\cal G})$ contains the bilinear Heisenberg couplings $\sum_a
\sigma_a^{(i)}\sigma_a^{(j)}$ enabling one to to implement swapping
between any pair of qubits \cite{divince}. Since controlled-{\sc not}
gates can be assembled as a sequence of ``square-root swaps'' and
single-qubit operations \cite{divince}, universal quantum logic can be
performed if we have access to fast modulation of single-qubit
Hamiltonians in addition to the required two-qubit interactions in the
centralizer.  A similar conclusion was conjectured in \cite{duan}.

Whether the proposed approach can be viable in realistic situations
will strongly depend on the details of the system and the
environmental noise, as well as on the sophistication of the available
technology. Bang-bang control is potentially suitable for NMR quantum
computation, as long as b.b. pulses are able to be effected with a
bandwidth small on the scale of the required spectral resolutions. 
At present, an experimental demonstration of b.b. control has been
reported for all-optical quantum circuits \cite{kwiat}. Despite the
challenges involved, the appeal of limited space resources may
stimulate efforts to practically implement decoupling in different
quantum information processors.

{\it Conclusion.$-$} We showed how to achieve noise-tolerant universal
quantum control on the full Hilbert space of the system based on
purely unitary open-loop manipulations. From the perspective of
quantum information processing, this implies the potential of
accomplishing noise-protected universal quantum computation without
the cost of extra space resources.  The method, which is best suited
for slow-response non-Markovian quantum baths, complements existing
approaches based on quantum error correction, where typically a
memoryless error scenario is assumed. As a further step toward the
goal of a truly fault-tolerant computation scheme, the present
assumption of perfect control resources should be relaxed to allow
{\sl noisy} decoupling and programming operations. A separate analysis
of the issue of robustness shall be presented in a future work.

This work was supported in part by DARPA/ARO under the QUIC initiative.
E. K. received support from the Department of Energy, under contract 
W-7405-ENG-36, and from the NSA. L. V. acknowledges partial support
from NSF-PHY-9752688.

${}^\dagger$ vlorenza@mit.edu; slloyd@mit.edu; knill@lanl.gov

\end{multicols}
\end{document}